\documentclass[12pt]{amsart}
\usepackage{a4wide,palatino,epsfig,latexsym,natbib,amsmath,amsthm,graphicx,
hyperref,enumerate}

\title[Antibiotic resistance landscapes]
{Antibiotic resistance landscapes:
a quantification of 
theory-data incompatibility
for fitness landscapes}

\author[Crona et. al]{Kristina Crona, Dayonna Patterson, Kelly Stack,
Devin Greene, Christiane Goulart, Mentar Mahmudi, Stephen D. Jacobs, 
Marcelo Kallman, Miriam Barlow}

\email{kcrona@ucmerced.edu}
        
\theoremstyle{plain}

\newtheorem{theorem}{Theorem}[section]

\theoremstyle{definition}

\newtheorem{definition}[theorem]{Definition}

\newtheorem{remark}[theorem]{Remark}

\newtheorem{example}[theorem]{Example}

\begin{document}

\begin{abstract}
Fitness landscapes are central in analyzing evolution, 
in particular for drug resistance mutations for 
bacteria and virus. We show that the fitness
landscapes associated with antibiotic resistance  
are not compatible with any of the classical models;  
additive, uncorrelated and block fitness landscapes.
The NK model is also discussed.

It is frequently stated that
virtually nothing is known about fitness landscapes
in nature. We demonstrate that available 
records of antimicrobial drug mutations 
can reveal interesting properties of fitness landscapes
in general. We apply the methods  
to analyze the TEM family of $\beta$-lactamases 
associated with antibiotic resistance.
Laboratory results agree with our observations. 
The qualitative tools we suggest are
well suited for comparisons of empirical 
fitness landscapes. Fitness landscapes are central in the
theory of recombination and there is a potential for 
finding relations between the tools and 
recombination strategies.
\end{abstract}

\maketitle

\section{Background}

The fitness landscape
was introduced as a metaphor
for adaptation. Informally, the
surface of the landscape
consists of genotypes,
where similar genotypes are
close to each other,
and the fitness of a
genotype is represented as
a height coordinate. Adaptation can then
be pictured as an uphill
walk in the fitness landscape \cite{wr}.
It is frequently claimed that we know virtually nothing
about fitness landscapes in nature. Scarcity of fitness 
measurements along with the difficulty in measuring fitness,
are cited as reasons. The purpose of this work 
is to demonstrate that available records of drug resistance mutations 
can reveal interesting properties of the underlying fitness landscapes.
We suggest qualitative tools that are easy to apply and 
interpret in order to learn properties of fitness landscapes from data. 
The setting we have in mind is
a record of clinically found 
antimicrobial drug resistance mutations, where there is a well 
defined wild-type and several mutant variants with 
some degree of drug resistance. 
Other cases of adaptation could work as well.

There are several advantages with mutation records
as a source of information. The records are 
already available. The quantity of data
is substantial and growing, and the quality tend to be high since 
the data is of medical importance. 
Moreover, the data reflects nature,
whereas laboratory data sometimes
disagree with clinical observations, 
see our discussion about
the TEM-family below.
However, the most important
reason is that this
data reflects adaptation.
In contrast, many times
empirical studies
where fitness is measured
consider combinations of
deleterious mutations.
The majority of such 
combinations are probably 
exceedingly rare in nature. If 
one is interested in adaptation,
one needs knowledge about 
beneficial mutations as well. 
For an overview of
recent empirical work
where fitness is measured,  
see e.g. \citet{ssf, wwl, ch}
and references.
Most existing studies concern
few loci (4 or 5)
and a specific
selective  environmnet.
For studies with many loci, 
see e.g. \citet{klh, ssk},
and for a case where fitness
ranks for the same genotypes
are determined in several different
selective environments,  
see \citet{gmc}.
 
We apply our results to the TEM family of beta-lactamases 
associated with antibiotic resistance. TEM stands for Temoneira, 
the name of the patient from whom the enzyme was first isolated.
TEM beta-lactamases have been found in {\emph{Escherichia coli}},
{\emph{Klebsiella pneumoniae}} and other gram-negative bacteria.
TEM-1 is considered the wild-type. The length
of TEM-1 is 287, i.e., 
TEM-1 can be represented as a sequence of
287 letters in the 20-letter 
alphabeth corresponding
to the amino acids.
Over 170 TEM variants have been found clinically, where 41 
are single mutants, i.e., they have
exactly one amino acid substitution,
and the the majority (90 \%)
have at most 4 amino acid substitutions. 
We use the record of the TEM family 
from the Lahey Clinic 

\url{http://www.lahey.org/Studies/temtable.asp}.

The quality of the TEM
record (from now on we will simply refer to 
"the TEM record"), is assumed to be high  
The TEM record represent a case of
multiple environments, but probably
not an exessive amount of completely different
environments. Several antibiotics have similar
effects. We have good reason to believe
that the TEM record is fairly complete,
and that there is not an abundance of
neutral mutations.

As a complement, we use laboratory results \citep{gmc}. Our study 
determines if TEM data is compatible with classical models of 
fitness landscapes. More precisely, we compare with 
additive, uncorrelated and block models of fitness landscapes. 
The NK model is also dicussed.

Throughout our article, our focus is
to what extent beneficial mutations combine 
well. We start with a brief description of our approach in the
context of the TEM family. 
Using standard notation, TEM-2 is a single mutant 
with the mutation Q39K, which means that the amino acid 
denoted "Q" (glutamine) at position 39 of the wild-type 
is substituted by the amino acid denoted "K" (lysine). 
TEM-174 is the single mutant A213V. 
One can ask if the double mutant with 
substitutions Q39K and A213V confer antibiotic resistance, 
since the double mutant combines two resistance mutations. 
However, the double mutant does not occur in the record.

Roughly, we compare the candidates 
for double mutants, such as the one described, with double
mutants that do occur in the record,
and consider the patterns for how 
candidates occur of are absent in the record.
This approach is motivated by an evolutionary 
perspective. Provided that the quality of a
record of resistance mutations
is good, most single mutants are more
fit, i.e., confer more drug resistance than
the  wild-type in som environment. Likewise, if a
double mutant occur in a mutation record, it is
plausible that the double mutant
is more fit than at least one of the corresponding
single mutants in some environment.

Put briefly, we    
consider if "good+good=better" for mutations.
The goal is to
capture the relation between 
fitness landscapes and
mutation records. 
One reason for being interested in 
mutation records, is that
laboratory results do not
always reflect clinical 
facts. A striking example
is that the triple mutant
of the TEM family
with substitutions A42G, E104K, and G238S,
confer a high degree of cefotaxime resistance
according to laboratory results \cite{wdd}, but this
mutant has never been 
observed clinically.
Examples where single 
mutants not found outside of the laboratory
confer a high degree of cefotaxime resistance
are given in \cite{ssk}.

It may seem surprising qualitative information
can be useful for analyzing fitness landscapes. However, we will
show that predictions from some classical models  
relate well to qualitative
information. Our approach
could be useful for comparisons of empirical 
landscapes, and there is a potential for 
relating the information derived 
directly to recombination strategies. 

We define fitness as the expected 
reproductive success, and use the
convention that the wild-type has 
fitness 1. Fitness is called {\emph{additive}} 
if the fitness effects of mutations
sum. Consider a biallelic two-loci system.
Suppose that the genotype ab has fitness 1,
the genotype Ab has fitness 1.03 and the genotype aB has 
fitness 1.01. If fitness is additive,
then the genotype AB has fitness $1.04=1+0.01+0.03$
(In the literature non-epistatic fitness is sometimes
defined as multiplicative, so that
the double mutant would have fitness 1.0403.
If the fitness effects and the number of beneficial mutations are small,
there is not much difference between the definitions.)
Values greater than 1.04 implies
positive epistasis. Values smaller than 1.04 implies 
negative epistasis. 

{\emph{Sign epistasis}} means that a particular 
mutation is beneficial or 
deleterious depending on 
genetic background. For example, if ab, Ab, aB
and have fitness values as above (1, 1.03, 1.01), but
AB has fitness 1.02, 
then there is sign 
epistasis. Indeed, in this 
case the mutation  B is beneficial 
for the ab-genotype, and 
deleterious for the Ab-genotype. 

The concept of a fitness 
landscapes has been formalized in 
different ways. A genotype  may be
represented as a string in the
20, 4 or 2 letter alphabet, 
depending on if one considers the
amino acids, the base pairs or
biallelic systems.
Thorughout the paper,
we will consider amino acids.

Let $\Sigma$ denote the 20 letter alphabeth.
The {\emph{genotype space}} $\Sigma^L$ consists of all
$20^L$ strings of length $L$.
A {\emph{fitness landscape}}
$w: \Sigma^L \mapsto \mathbb{R}$ assigns
a fitness value to each genotype.
The fitness of a genotype $g$
is denoted $w_g$.
If two genotypes differ by
a single mutation, they are
{\emph{mutational neighbors}}. 

\begin{remark}\label{og}
Following the Orr-Gillespie approach we 
assume that the wild-type has very high 
fitness also in the new environment as 
compared to a randomly generated genotype.
Consequently, only a small number of mutations
of the wild-type are beneficial. 
\end{remark}

The paper is structured
as follows. In Section 1.1 we briefly review
classical models of fitness landscapes.
Section 2 provides basic observations
of the TEM record.
Section 3 concerns
additive and uncorrelated fitness,
and Section 4 block models.
For all models, we compare with the
TEM record, and in Section 5
a laborator study of TEM alleles
is used as a complement to the record.

\subsection{Classical models of fitness landscapes}
Additive fitness landscapes, uncorrelated fitness landscapes,
the block model and Kauffman's NK model
have had a broad influence in evolutionary biology. We will 
give a brief overview of the four classical models.

Additive fitness landscapes, 
or non-epistatic landscape, has been defined.
An additive fitness landscape
is single peaked.

In contrast, for an uncorrelated (also called random, rugged
or House of Cards [HOC]) fitness landscape, there is no 
correlation between the fitness of a genotype and 
the fitness of its mutational neighbors, i.e., alleles that 
differ by one substitution only. 

Consider an uncorrelated landscapes where say $2\%$ of the
single mutants are more fit than the wild-type. 
It follows that for double mutants corresponding to two beneficial
single mutations, approximately $2\%$ are more fit than the 
wild-type as well. In other words, beneficial mutations do usually not
combine well for an uncorrelated fitness landscape by Remark \ref{og}.

Uncorrelated fitness and additivity can be considered as 
two extremes with regard to the amount of structure 
in the fitness landscape, and most fitness 
landscape fall between the extremes.
Uncorrelated fitness has been studied extensively in the 
literature \citep[see e.g.][]{k, kl, fl, rbj, pk}.

For the block model (see \citep[e.g.]{mp, o2006})
the string representing a genotype can be subdivided into blocks, 
where each block makes an independent contribution
to the fitness of the string. Each block has uncorrelated fitness, 
and the fitness of the string is the sum
of contributions from each block.
In particular, a block model
consisting of only one block only is
an uncorrelated fitness landscape.
The rational behind this model is
that if two blocks have completely different functions, 
then the effect of two changes in 
different blocks should be independent.

Kaufmann's NK model (see e. g. \cite{kw})
is defined so that the
epistatic effects are random, whereas the fitness of a genotype is
the average of the "contributions" from each locus.

More precisely, for the NK model
the genotypes have length N (in our notation $L=N$),
and the parameter $K$, where $0 \leq K \leq N-1$,
reflects interactions between loci.
The fitness contribution $\phi_i$  from the locus $i$
is determined by its state $g_i$ and the states at $K$ 
other loci $i_1, \dots, i_K$. 
The key 
assumption is that this contribution
is assigned at random from some
probability distribution.  
The fitness of a genotype $g$
is the average of the contributions $\phi_i$,
so that 
\[
w_g=\frac{1}{N} \sum_{i=1}^{N} \phi_i (g_i, g_{i_1}, \dots, g_{i_K}),
\]
where $i_1, \dots, i_k \subset \{1, \dots, i-1, i+1, \dots, N \}$.
Several important properties of NK landscapes depend 
mainly on N and K, rather than the exact 
structure of the epistatic interactions.
The fact that the fitness of the genotype is
the average of these $N$ contributions, means
that fitness effects of non-interacting mutations sum.
Notice that the fitness landscapes is additive
for $K=0$ and uncorrelated for $K=N-1$.
The popularity of the NK model
rests on the that
the model is "tunably rugged".
This expression means that 
the ruggedness is expected to increase
by $K$ from the single peaked additive
landscape for $K=0$ to the
uncorrelated landscape with a maximal
number of peaks for $K=N-1$.
Published results on the NK model
of (potential) relevance to evolutionary biology
concerns the number of peaks,
the length of mutational trajectories,
fitness distributions of genotypes
and fitness trajectories.

Notice that also the block model includes 
additive landscapes and uncorrelated landscapes as special cases. 
More importantly, NK models
and block models  are similar in that there is a 
sharp division between effects which are completely random 
and effects which are additive.
One should keep in mind that
the block models and Kaufman's NK model
are equipped with very special structures. 
In order to provide some intuition
for how empirical landscapes relate
to the models, we will consider
examples.

\begin{example}
\[
w_{000}=1, 
w_{001}=w_{010}=w_{001}=1.01,
w_{011}=1.02, w_{101}= w_{11}=1.02, 
w_{111}=1.03
\]
For every loci, replacing 0 by 1 increases fitness
by $0.01$. Fitness is additive and the genotype $111$ is at a peak.
\end{example}

\begin{example}
\begin{align*}
&w_{0000}=1, w_{0001}=w_{0010}=w_{0100}=w_{0001}=1.02, \\ 
&w_{0011}=w_{0101}=w_{0110}=1.03, w_{0111}=1.035, \\
&w_{1000}=1.02, w_{1001}=w_{1010}=w_{1100}=1.04, \\
&w_{1011}=w_{1101}=w_{1110}=1.05,  w_{1111}=1.55.
\end{align*}
For every loci, replacing 0 by 1 increases fitness.
For the first locus the increase is  $0.02$, 
regardless of background.
For the other loci, the magnitude of the difference depends
on the background. For instance, the magnitude is
0.02 for the change from 0000 to 0001,  0.01 
for the change from 0001 to 0011, and 
0.005 for the change from 0011 to 0111.
\end{example}

Fitness is obviously not additive in the second
example, since the fitness of the double and triple mutants
are below linear expectations based on the wild-type
and the single mutants. The landscape deviates from
expectations for an uncorrelated landscape
as well, since replacing 0 by 1 always gives higher
fitness. 
It remains to consider the more
general models. 
As for the block model, the 
first locus is independent,
The remaining three loci 
interact with each other
in a symmetric way.
Consequently, the natural candidates
for blocks would be one block
consisting of the first locus,
and another block consisting
of the remaining three loci.
However, the second block
deviates considerably from random expectations.
Consequently, the block model is not a good fit.

For the NK model,
the independent fitness contribution
for the first loci suggest that $K=0$.
However, the second locus depends on
the third and the fourth loci, suggesting
that $K=2$. (Similar arguments
for the third and the fourth loci,
suggests that $K=2$ as well.) 
Since the observations suggest
different $K$-values,
the NK model does not seem
ideal.

\begin{remark}\label{sym}
The NK model allows different
interactive patterns. However,
it is not expected that
some loci are more or less independent,
whereas other loci have considerable
interactions. Some degree of
of symmetry is expected, reflecting
the $K$ value. 
\end{remark}

Several other models for fitness
landscapes have been suggested,
including  neutral models, see 
\cite{ssf} for an empirical perspective.
For some approaches to fitness landscapes
not related to the models mentioned, 
see the geometric theory of
gene interactions \cite{bps,c},
and the Orr-Gillespie theory
\cite{o2002}.
Notice also that fitness landscapes have been
used in chemistry, physics
and computer science, in addition to evolutionary biology.
In combinatorial optimization the fitness function
is referred to as the cost function. For a survey
on combinatorial landscapes in general see \cite{rs}.

\section{The qualitative measure of additivity and the TEM record}
Throughout the paper, we focus
on single and double mutants in a record.
The motivation is trivial. If a single mutant has high fitness,
it is likely to be found in nature. However, if
a $k$-tuple mutant is very fit for some large $k$, the mutant
may never be found because of the time
span necessary before the substitutions have accumulated. Single and
double mutants are likely to appear relatively early in the process of
adaptation.
Roughly, we are interested in the proportion of beneficial
mutations among all possible single mutations, as well
as to what extent beneficial mutations
combine well.
The information we consider is coarse,
and a record of mutation will rarely be perfect.

As indicated, we will work with words in the 20 
letter alphabet where there is a well defined
wild-type. A single mutant
is a genotype resulting from one amino acid
substitution.
However, the amino acid substitutions
are not comparable 
to substitutions of
letters in a string.
Not every amino acid substitution   
can occur as the result 
of a single point mutation.
For instance, suppose that the 
amino acid is Valine at 
a particular locus, and that
the codon is GTT. 
Then one can obtain
exactly 6 single mutants
at the locus corresponding to A, D, F, G, I,
L (or Alanin, Aspartic acid,  Phenylalanine, Glysine  
Isoleucine, Leucine).
On the other hand,
one can obtain
exactly 8 single mutants (A, D, E, F, G, I, L, M)    
starting from Valine (the 
codons for Valine are GTT, GTC, GTA, GTG).

In general, the number of single mutants 
one can obtain varies depending on amino acid. 
Moreover, in some cases, such as for Valine, the 
number depends on if one consider
a particular codon for the 
amino acid or all possible
codons. 

To make matters more complicated, 
the wild-type allele under consideration
may be unique in terms of amino acids, 
but not in terms of codons. For a precise
analysis, one may want to consider
codon variations in the wild-type allele.
However, for our purposes it is sufficient
to consider amino acids.

We assume that there are
approximately 7 possible single point 
mutations for a given locus, so
that if the wild-type has length 
$N$, there are $6N$ mutational neighbors.
For the reader's convenience, we included a
table of possible single mutants (see Section 7).

\begin{remark}\label{sex}
Throughout the paper, we assume a genotype
of length $N$ has $6N$ mutational neighbors.
\end{remark}

We are interested in the proportion of beneficial
mutations among all possible mutations.
By Remark \ref{og}, the proportion 
of beneficial mutations is expected to 
be small.
For the TEM record $N=287$ and there are 47 single
mutants in the record. Consequently, 
\[
\frac{S_R}{6N}=\frac{6}{6 \cdot 287}=2.67 \%
\]

Another interesting property of a fitness landscape,
is how beneficial mutations combine.
More precisely, consider a double mutant which
combines two beneficial single mutations. 
If the double mutant is less fit than both
single mutants, then the double mutant would most likely 
not appear in the record. The qualitative measure of additivity
is motivated by this observation. More precisely, 
we will use the following definition.

\begin{definition}
Let $B_p$ be the set
consisting of all double mutants such that both
corresponding single mutations are beneficial.
The set $B \subset B_p$ consists of all double 
mutants in $B_p$ which are more fit than at
least one of the corresponding single
mutants. The qualitative
measure of additivity for a fitness landscape is the ratio
$\frac{|B|}{|B_p|}$.
\end{definition}

Consider the single mutations in a record and the corresponding double        
mutants. Whenever two single mutants at different sites occur in the
record, the corresponding
double mutant is considered a
candidate for a double mutant of high fitness.
Let $\hat{B_p}$ be
the set of candidates for double mutants. 
Let $\hat{B} \subset \hat{B_p}$ 
be the set of double 
mutants in the record among the candidates.
Loosely speaking, one can consider
$\frac{|\hat{B}|}{|\hat{B_p}|}$
the observed qualitative measure
of additivity. Under ideal circumstances, the
ratios $\frac{|\hat{B}|}{|\hat{B_p}|}$
and $\frac{|B|}{|B_p|}$
are approximately the same, at least in
for antimicrobial resistance mutations
in the context we consider.
We assume that the adaptation, or the
resistance development,  will take place
repeatedly at different geographic
locations. If a double mutant is more
fit than at least one of the single
mutants, the double mutant should
occur sooner or later.

If fitness is additive, then
$\frac{|B|}{|B_p|}$=1, and
for uncorrelated fitness one
expects the value 
be close to 0 by Remark \ref{og}.
Of course the measure is coarse.
However, it is valuable to
have a simple method for comparing
fitness landscapes in different contexts. 
Whenever fitness is measured,
one can determine
$\frac{|B|}{|B_p|}$, 
and for any record one
can determine  $\frac{|\hat{B}|}{|\hat{B_p}|}$.
Notice that
one expects the qualitative measure
to decrease by increasing block size 
for the block model, as well as  
by increasing $K$ for the NK model.

For some background,
a measure of additivity
which reflects quantitative fitness
differences is called "roughness" \citep{ch, aih}.
Roughness 0 implies that the landscape
is additive. A problem with roughness 
is a possible size bias,
i.e., all else equal, the
roughness may 
be greater for a large
number of loci.
The qualitative measure of additivity
does not have any size bias.

Analyzing epistasis is closely
related to analyzing additivity.
For a thorough discussion about different  
measures of epistasis and empirical 
fitness landscapes, see \cite{ssf}. 
The most fine-scaled approach to epistasis
is the geometric theory of gene
interactions, which 
uses triangulations of polytopes
\citep{bps,bpsb, c}.

We will consider the 
$\frac{|\hat{B}|}{|\hat{B_p}|}$ value
for the TEM record.
The record has 46 single mutants.
The substitutions are at position
position 69 for 3 single mutants,
at position 164 for 3 single mutants,
at position 244 for 5 single mutants
and at position 275 for 2 single
mutants. Each remaining single mutant
has its mutation at a unique position.

It follows that the number of candidates are
\[
{46\choose 2}-
{3\choose 2 }-{3\choose2 }-{5\choose 2}-{2\choose 2}
=1018.
\]

The record has 35 double mutants in the set $\hat{B}$
(see Table 1 for a list of the double mutants in the set $\hat{B}$).
Consequently, 
\[
\frac{|\hat{B}|}{|\hat{B_p}|}=\frac{35}{1018}= 3.43 \% 
\]

We summarize the results for the TEM record in the
following observation.

\medskip
{\bf{Observation 1.}}
For the TEM record,
\begin{enumerate}
\item the proportion of beneficial single mutations 
is $2.67 \%$,
\item the ratio
$
\frac{|\hat{B}|}{|\hat{B_p}|}=\frac{35}{1018}= 3.43 \%. 
$
Consequently, $3.43 \%$ is an estimate
of the qualitative measure of 
additivity $\frac{|{B}|}{|{B_p}|}$.
\end{enumerate}

\section{records of mutations, additive fitness and uncorrelated fitness}
Consider a record of drug resistance mutations.
We first consider conditions ideal for our
purposes. Then we discuss the consequences of
relaxing some of the conditions.

\subsection{The perfect record conditions.}
Assume that we have a well defined wild-type and several mutant variants
associated with drug resistance. Assume that the records of drug
resistance mutations
satisfy the following conditions.
\begin{enumerate}
\item The organism adapts to a single
environment. [single environment condition]
\item The record is complete with respect
to single and
double mutants in the sense that
\begin{enumerate}
\item
All single mutants which are more fit than the wild-type occur in the record,
\item
All double mutants which are more fit than both corresponding single
mutants occur in the record. [completeness condition]
\end{enumerate}
\item 
Every single and double mutant in the record is a result of
adaption. In particular, the single mutants are the result
of beneficial mutations. [Absence of neutral mutations condition] 
\end{enumerate}

\begin{remark}\label{sats}
Assume that a record  satisfies the
perfect record conditions, as described.
\begin{enumerate}
\item[(i)]
If $\frac{|\hat B|}{|\hat B_p|}<1$, then the fitness landscape is not 
additive
\item[(ii)]
Suppose that there are $s_R$ single mutants in the record. If the fitness 
landscape is uncorrelated, then one 
expects
$
\frac{|\hat B|}{|\hat B_p|} 
$
to equal
$
\frac{s_R}{9L}
$
under the (simplified) assumption that a genotype has
$6N$ mutational neighbors.
\end{enumerate}
\end{remark}
The first claim is obvious. Fitness being uncorrelated, 
approximately
$\frac{s_R}{6L}$ of double mutants are more fit than the wild-type. If one 
restricts to the
category of double mutants where both corresponding single mutants are 
more
fit than the wild-type, the proportion is $\frac{s_R}{6L}$ as well. 
Fitness being uncorrelated,
one third of the double mutants in this category are expected to be less 
fit than both single mutants. Indeed, there
are three possible 
fitness ranks, and the double mutant is as likely to
have the lowest fitness as any other rank. 
The resulting proportion
is
\[
\frac{2}{3} \cdot \frac{s_R}{6L}=\frac{s_R}{9L},
\]
which explains the second claim in the remark.

\subsection{Relaxing the perfect record assumptions}
The TEM family has adapted to
different selective environments,
since antibiotics have different
effects, so that the single environment
condition is not satisfied. 
First we relax the single environment condition
for a record.

\noindent
{\emph{Multiple environments and additive landscapes.}}
In contrast to the single
environment case, even if the fitness landscape associated
with each drug is additive the 
$\frac{|\hat{B}|}{|\hat{B_p}|}$-value may be lower  than 1.
The reason is that if two different single mutants
are adapted to different environment, then a combination
of the two corresponding mutations may not be fit in any 
environment. As an illustration, consider 
the following examples with
additive landscapes.

\begin{example}
Consider two different environments and  50 single
resistance mutations, where 25 mutants are adapted
to each one of two environments.
Assume that the fitness landscapes associated
with both environments are additive. Moreover, assume that
two mutations that constitute adaptations to different
environments never combine well, so that
the corresponding double mutants do not occur
in the record.
Then
\[
\frac{|B|}{|B_p|}=\frac{ {25 \choose 2} + {25 \choose 2}}{{50 
\choose2}}
=0.49 .
\]
Consider exactly the same situation with 50
single mutants but instead 10 different environments,
where 5 single mutants are adapted to each different environment.
Then
$
\frac{|B|}{|B_p|}=\frac{ 10 \cdot {5 \choose 2}}{{50 \choose2}}=0.082.
$
We conclude that in the case of multiple environments the
$\frac{|B|}{|B_p|}$-value may be low even if each fitness landscape is
additive.
\end{example}

The case described, where
mutations which are beneficial in different
environments {\emph{never}} combine well
is probably not realistic. However, it is clear that
multiple environments may lead to a 
lower $\frac{|B|}{|B_p|}$-value.

\noindent
{\emph{Multiple environments and uncorrelated landscapes}}.
Consider a situation with multiple
environments where the fitness landscape associated
with each environment is uncorrelated. For simplicity,
we assume that there are not an excessive amount of
different environments.
By assumption, very few 
single and double mutants should be 
more fit than the wild-type 
in any particular single 
environment. Multiple environments
imply more chance for a mutant
to be fit in at least one environment.
However, fitness being uncorrelated,
that effect is exactly the same for 
single and double mutants.

Double mutants will be more fit than the
wild-type in any of the different environments,
so that the  $\frac{B}{B_p}$-value will be very low also
in the case of multiple environments. In other words, 
unless the $\frac{B}{B_p}$-value is very small, we can rule
out that all landscapes are uncorrelated fitness landscapes.
(Multiple environments may lead
to more beneficial mutations. However, there
is no difference between single and double mutants 
in that respect, so that the $\frac{B}{B_p}$-value should
not be influenced.)

\noindent
{\emph{Incomplete records.}}
Missing single mutants in the record
will normally have little effect, since
$\frac{|\hat{B}|}{|\hat{B_p}|}$ 
concerns only single mutants in the record and associated
double mutants, by definition. However, missing double
mutants will make the $\frac{\hat{B}}{\hat{B_p}}$-value
smaller as compared to the result for a more 
Consequently, incompleteness may lead to
and underestimate of $\frac{|B|}{|{B_p|}}$.

\noindent
{\emph{Neutral mutations.}}
An abundance of neutral mutations
will make the 
$\frac{|\hat{B}|}{\hat{|B_p|}}$-value
difficult to interpret.

\begin{remark}
An abundance of neutral mutations
make the record difficult to interpret.
In the case of multiple environments or
incomplete records, Observation 2 (i) holds, 
but not 2 (ii).
\end{remark}

The TEM record represents a case of
multiple environments, but probably
not an excessive amount of completely different
environments. We have good reason to believe
that the record is fairly complete,
and that there is not an abundance of
neutral mutations in the record.

For the TEM-record 
$
\frac{\hat{B}}{|\hat{B_p}|}=\frac{35}{1018}= 3.43 \%,
$
from Observation 1,
and 
\[
\frac{s_R}{9L}=
\frac{46}{9 \cdot 287}=1.78 \%.
\]

{\bf{Observation 2.}}
\begin{enumerate}
\item[(i)] Under the perfect record assumptions, 
the TEM landscape is not compatible with
additive or uncorrelated 
fitness landscapes.
\item[(ii)] The TEM record is not compatible with
uncorrelated fitness landscapes
under realistic assumptions for the TEM record.
\item [(iii)]
The TEM record combined with
knowledge of the context, 
suggest that fitness is
not additive for the TEM family. 
\end{enumerate}
Part (iii) rests on the 
fact that there does not exist an
excessive amount of
completely different environments
for the TEM family.
It would be remarkable with 
$
\frac{\hat{B}}{|\hat{B_p}|}= 3 \% 
$
if all the fitness landscapes
associated with individual  drugs
were additive.

The TEM record has in 
total 46 double mutants,
where 35 are included in 
the set $\hat{B}$.
We conclude the section with some
remarks about the remaining double
mutants (see Section 7 for a list of 
them, and Table 2 of the same section
for a list of the double mutants in $\hat{B}$).
For the double mutant
TEM-164, none of the single substitutions correspond
to single mutants in the record.
For the other 9 double mutants,
exactly one of the substitutions
corresponds to a single mutant.
The most likely reason for 
a double mutant in $B_p$
not to be included in $B$ 
is sign epistasis. Specifically,
the single mutation not in the record
is selected for only if the 
other single mutation has already occurred.
In such a case, 
the \emph{sign} of the effect (positive
or negative) of the second mutation depends
depends on background (the 
effect is negative for the
wild-type and positive if the
first mutation has occurred).
Constraints for orders in which mutations accumulate
are known from different contexts
\citep[see e.g.][]{djk, bes}, including HIV drug
resistance. 

\section{Records of mutations, block models and position graphs}
If fitness is neither additive nor uncorrelated, 
then one may consider
more general models.
We will discuss block model
with focus on how single
beneficial mutations combine.
However, in this context one
has to consider the 
structure for
how beneficial mutations
combine, not only 
the proportion of
good combinations.
For simplicity,
we will discuss loci rather than
amino acid substitutions.
The position graph
is intended to display
the structure of the combinations.

\begin{definition}  
For a record of mutations,
each node of the
{\emph{position graph}}
corresponds to
a locus associated
with a single mutant
in the record. An edge between two
nodes indicates that a double mutant
occurs in the record,
such that the two mutations 
correspond to the nodes.
\end{definition}

Notice that 
the position
graph reflects
the sites but ignores
the actual amino acid
substitutions (such as
if the substitution is
glutamine or lysine, or
if both of them occurs at the site).
Single mutants with substitutions
at the same site may of
course differ    
in how well they combine
with other mutation.
One may want to look
at more-fine scaled information
and distinguish between different
substitutions at the same site.
However, for simplicity, we
ignore this complication.

Figure 1 shows the 
position graph for the TEM family,
except that nodes of degree
zero are omitted.

\begin{remark}
The position graphs considers
loci, but not the amino acid
substitution. One may want to look   
at more-fine scaled information
and distinguish between different 
substitutions at the same site.
\end{remark}

\begin{figure}
\includegraphics[scale=0.5]{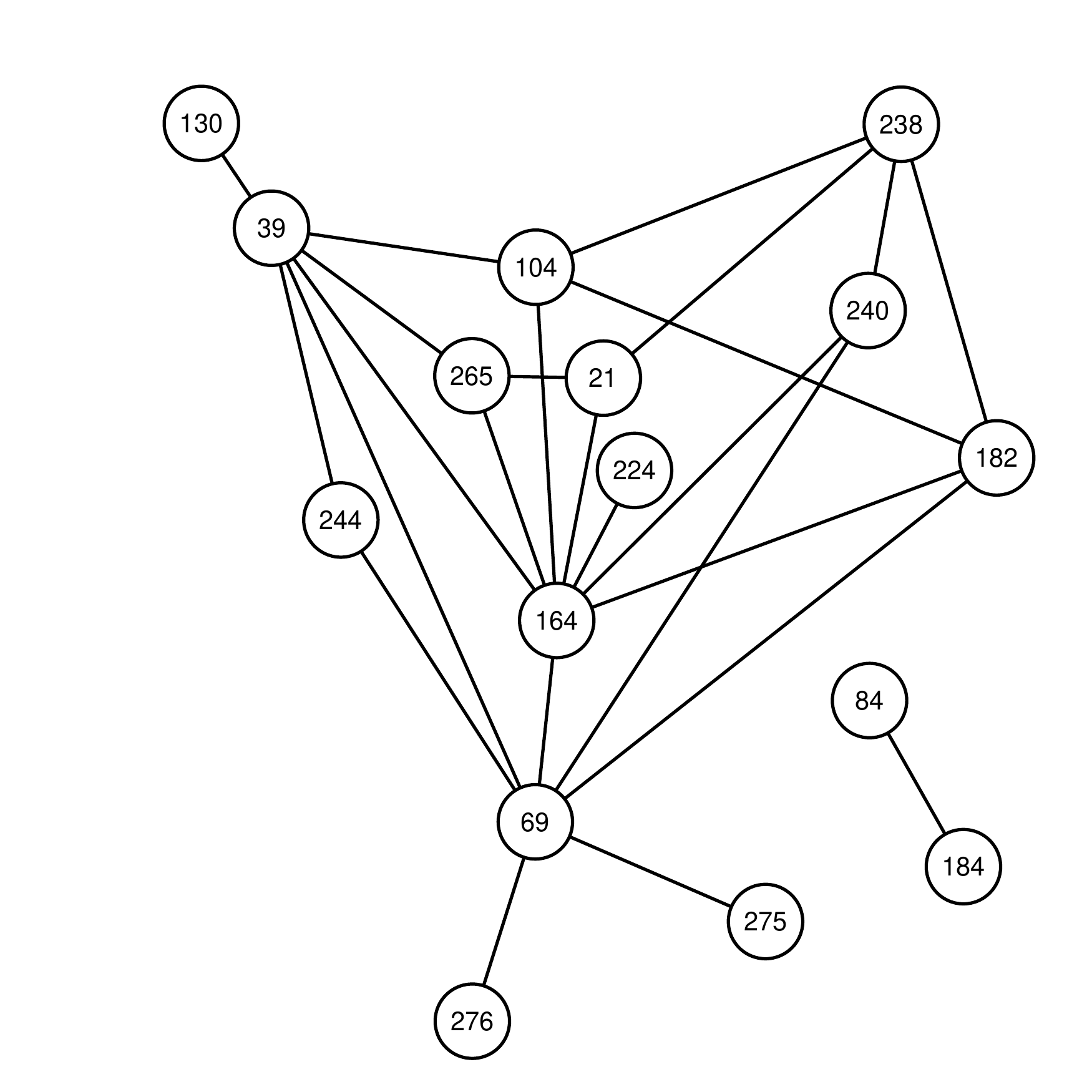}
\caption{The position graph for the TEM-family,
where we have omitted
the  21 nodes of degree 0.}
\end{figure}

Recall that the
{\emph{degree}}
of a node is the number of
edges to other nodes.
The complement $\overline{G}$ of a
graph $G$ is a graph on the same nodes,
where a pair of nodes are connected
by an edge exactly if the pair
is not connected by an edge for
$G$.
For a {\emph{complete bipartite graph}},
the nodes can be partitioned
into two subsets, such that
every pair of nodes 
from different subsets 
are connected by an edge,
and there are no other edges.

The following observation is elementary
by Remark \ref{og}.
\begin{remark}
Assume the block model (with at least two blocks).
Let $G$ denote the position graph.
Consider $G$ and the complement
$\overline{G}$. Under the
perfect record assumptions,
the nodes of $G$ have degree one or more. 
Moreover, by Remark \ref{og},
the following statements hold
modulo a few errors:
\begin{enumerate}
\item For the case of two blocks, $G$ is
a complete bipartite graph. 
It follows that  $\overline{G}$ is a disconnected
graph with two components, both
of which are complete.
\item
In general, for $l$ blocks,
$\overline{G}$ is a disconnected
graph with $l$ components, all of
which are complete.
\end{enumerate}
\end{remark}

For the TEM record,
the single mutants
correspond to substitutions at
exactly 37 positions. 
Exactly 21 nodes out of the 37
have degree zero.
The position graph has 
37 nodes and 25 edges.

Consider the block model (for at least two blocks). 
The following observations
are potentially problematic 
for a block model.

\begin{enumerate} 
\item There is an abundance of nodes of degree zero (21 our of 37), 
\item the total number of edges is (only) 25 and there  are 37 
nodes.
\item the maximal degree for a nodes is 8, 
\item $G$ has several triangles,
in particular a triangle consisting of
the three nodes of highest degree 
out of all nodes of $G$. 
\end{enumerate}

Under the perfect record assumptions,
the block model implies that the degree
of each node is at least one (provided that
the nodes are distributed over at least two 
blocks).

For the case of two blocks, the  number of edges is
between 36 and $18 \times 19=342$,
where the minimum corresponds to
the distribution 1 and 36 nodes per block, and
the maximum corresponds to the distribution
of 18 and 19 nodes per block.
In the first case, one node should have degree 36.
However, the maximal degree of nodes in 
the position graph (Fig. 1) is 8.
It is clear that the position graph has too 
few edges for similar
block lengths, and too low maximal degree
for unequal block length.

For more than two blocks, 
even more edges are expected
leading to similar problems.
Clearly the block model
is not compatible with the
data under the perfect record
conditions.

Consider the case of exactly two blocks
in a more realistic situation.
Then the position graphs should have
essentially no triangles by Remark \ref{og}. This is because
at least two of the nodes in a triangle 
are on the same block.
(Moreover, for two blocks a reasonable
guess would be that the three nodes of highest 
degree (39, 69, 164) are on the same block [the shorter one].
If so, it is unexpected with a triangle consisting
of the three nodes.) 

It remains to consider
the single record condition
and consider more than two blocks.
In that case,
let us analyze the fact that
there are relatively few edges.
21 out of 37 nodes have 
degree zero, and
in fact one can 
find a set of 31 nodes (including the 21 nodes)
in the position graph, such
that {\emph{no pairs in the set are neighbors}}. 
That implies that among 31 nodes one 
cannot find a single pair of nodes
on different blocks, such that both of them
have high fitness in the same environment.

This is of course possible, especially
taken into account that the record
may be incomplete. However, from
knowledge of the context, it does not
seem plausible. The number of
completely different environments is
limited.

\medskip
{\bf{Observation 3.}}
From the TEM record 
and some knowledge of the context,
the  block model is probably 
not a good fit for the TEM family.

\section{A laboratory study of TEM alleles}
We will compare the results from
the TEM record with a laboratory study.
The advantage with the laboratory
study is that one can use
drug specific information, 
which takes care of the 
difficulties resulting from
multiple environments.

The study \citet{gmc}
considered the antibiotics
Ampicillin (AMP),
Ceftazidime (CAZ)
Cefpodoxime (CPD),
Cefprozil (CRP),
Cefotetan (CTT),
Cefotaxime (CTX),
Cefepime  (FEP)
and Pipercillin/tazobactam penicillin/inhibitor (TZP).
Fitness ranks were detremined for the 4 single mutants
L21F, R164S, T265M, E240K.
as well as the
6 double mutants than can be obtained
from them, for each of the 8 antibiotics.

In particular, for the drug
Ceftazidime (CAZ), 4 single mutants were more fit
than the wild-type, so  that
one can obtain 6 double mutants from
the single mutants. 
Out of the 6 double mutants
for Ceftazidime,
5 double mutants
were more fit than at least
one of the single mutants,
and one double mutant
(the combination
of R164S and E240K) 
was  more fit than both corresponding
single mutants.

For the 9 antibiotics, 
we list the number of single mutants
with higher fitness than the wild-typ.
and below the $\frac{|B|}{|B_p|}$-value.
\begin{align*}
&4, 4, 2, 2, 2, 2, 2, 2, 2 \\
&\frac{5}{6}, \frac{1}{3}, 1, 1, 1, 1, 0, 0, 0
\end{align*}
The mean values is 0.57.
Obviously the data deviates considerably
from additive fitness as well 
as uncorrelated fitness.

For a comparison,  combinations
of 5 beneficial mutations
from an experimental {\emph{Escherichia coli}} population
were considered in \citet{kds}.
Negative epistasis dominated, but sign epistasis was
rare. Consider the 10 double mutants combining pairs
of the 5 beneficial mutations. Every double mutant had higher
fitness than at least one of its corresponding single
mutant, so that $\frac{|B|}{|B_p|}=\frac{10}{10}=1$.

As for the block model, with very few exceptions,
the double mutants in $B$
should be more fit than {\emph{both}}
single mutants, or less fit than
{\emph{both}} corresponding single mutants,
by Remark \ref{og}.
For the 9 drugs, one can form in total
19 double mutants. 5 of them are more
fit than both corresponding single mutants,
6 of them are more fit than 
{\emph{exactly one}} corresponding single mutants,
and 8 of them are less fit than 
{\emph{both}} corresponding single mutants.
This observation indicates that the block
model does not apply. Notice also
that the most plausible block
distribution of nodes differ
from drug to drug (for some drugs
node 21 and 265 should
be on the same block, and for other
drugs not). 

\medskip
{\bf{Observation 4}}
Neither additive fitness,
uncorrelated fitness,
nor the block model 
is compatible with
data from the laboratory study 
\cite{gmc}.

\section{discussion}
We have compared expectations
from additive, uncorrelated,
and block models 
of fitness landscapes with
empirical data. 
We argue that the TEM family
of $\beta$-lactamases
is not compatible
with the three models.
Under the  simplified assumptions
of a complete record and a single environment,  
we found that the TEM data was not compatible with 
anyone of the three models.
Under more realistic assumptions
for the TEM family,
the data was not compatible with uncorrelated fitness. 
Similarly, using the record
and some knowledge of the biological context, 
it seems plausible that neither the
additive nor the block model
is a good fit. Our  conclusions for 
the three models
were confirmed by a laboratory study
of TEM alleles.
We did not compared the
TEM family and the NK model. 
However, the 
symmetry aspect (see Remark \ref{sym}) 
could be problematic.

The literature on
additive, uncorrelated
block and NK models literature 
is extensive.
Some approaches in the field 
have been motivated by theoretical
considerations, such as relating epistasis
to the number of peaks of a fitness landscape.
The purpose of our study was not to debate
the value of the classical models.
We appreciate that toy models can be used 
for generating fruitful hypotheses, 
which can be tested empirically.
As for additive and uncorrelated
fitness, the extremes will always
be of interest as a theoretical starting point.

However, the classical
models have been used for interpretations
of empirical data as well.
A standard assumption for several topics
in evolutionary biology and breeding is
additive (or multiplicative) fitness,
in particular for studies of fitness inheritance and 
sexual selection \citep[e.g][]{kbj}.
Statistical methods which are suitable
for uncorrelated fitness landscapes have
been used  in empirical studies 
\citep[see e.g.][for a discussion]{cgbb},
and the NK model is frequently used
in  empirical contexts.
From this perspective, it is reasonable
to discuss to what extent the
classical models are realistic.

We have suggested 
elementary tools, including the
qualitative measure of additivity
and the position graph, for
comparing models and data. 
Our approach demonstrates
that one can determine
properties of fitness
landscapes from a 
record of mutations.
The ideal setting is 
a single environment.
It would be of interest
to compare the qualitative
measure of additivity with
observed behavior for
microbes, such as recombination 
strategies. 
The position 
graph can be used as 
a test of modularity. 

We  consider it an advantage that
our approach does not depend on
any structural assumptions of
the underlying fitness landscapes.
Any case of adaptation where one has
a well-defined wild-type and
some direct or indirect
method for determining fitness ranks
of genotypes works.

Qualitative information has its limitations. The ideal information 
for determining properties of fitness landscapes is of course 
direct fitness measurements. For obvious reasons such information 
is sometimes difficult, if even possible, to derive. Moreover, 
laboratory results do not always agree with clinical findings. 
Consequently, direct methods for interpretations of nature 
are of interest as a complement to experimental results.

\bibliographystyle{apalike}

\newpage

\section{Statistics and tables}
We use information from
the record of the TEM family
from the Lahey Clinic
\url{http://www.lahey.org/Studies/temtable.asp}.
as of April 2012 for this study.
All mutants have been found clinically.
The record is continuously updated
with new mutants.

The following loci, in total 37, correspond to
single mutations:

21, 28, 34, 39, 68,
69, 84, 92, 104, 115,

120, 124, 130, 145, 155,
157, 158, 163, 164, 176,

182, 184, 189, 204, 213,
218, 224, 230, 238, 240,

244, 265, 271, 275, 276,
280, 283

\smallskip
Exactly 4 loci, out of
the 37 listed above,  correspond
to more than one substitutions
(see Table 1).
\medskip

\begin{table}
\caption{Loci with more than one substitution}
\centering
\begin{tabular}{l l l}
\hline \hline
locus & \# of substitutions \quad & alleles \\  
\hline
69:  &3 &TEM--33, TEM-34, TEM-40 \\
164: &3 &TEM--12, TEM-29, TEM-143 \\
244: &5 &TEM-30, TEM-31, TEM-51, TEM-54, TEM-79 \\
275: &2 &TEM-103, TEM-122 \\
\hline
\end{tabular}
\end{table}

\subsection{Double mutants and the position graph}
The total number of double mutants in the
record is  45, where 35 combine single mutations from the
record (see the table). The remaining 10 mutants are
as follows:
TEM-58, TEM-81, TEM-112, TEM-126, TEM-137,
TEM-145, TEM-146, TEM-163, TEM-164, TEM-169.

Information about double mutants in the record 
are expressed in the position graph (see Fig. 1).
Recall that for the position graph a node corresponds
to a locus with a single mutation. An edge
denotes that there exists at least one
double mutant which combines
the substitutions at the two loci.
The position graph has 37 nodes and 24 edges.
The following (21) nodes of the position graph
have degree zero.
\[
28, 34, 68, 92, 115, 
120, 124, 145, 155, 157, 
158, 163, 176, 189, 204, 
213, 218, 230, 271, 280, 
283
\]
The degree of the remaining (16) nodes
are listed by locus (locus:degree).
\begin{align*}
&21:3, \quad 39:6,\quad 69:7, \quad  84:1, \quad  104:3, \quad  130:1,  
\quad 164:8,  \quad 182:3,  \quad 184:1, \\ 
&238:4, \quad 224:1, \quad  240:3, \quad 244:2,  
\quad 265:3, \quad 275:1, \quad 276:1  \\
\end{align*}

\begin{table}
\caption{Single mutants of the record}
\centering
\begin{tabular}{l l l}
\hline \hline
1& TEM-2  & Q39K \\  
2& TEM-12 & R164S \\ 
3& TEM-17 & E104K \\ 
4& TEM-19 & G238S \\ 
5& TEM-29 & R164H \\
6& TEM-30 & R244S \\ 
7& TEM-31 & R244C \\    
8& TEM-33 & M69L \\ 
9& TEM-34 & M69V \\  
10& TEM-40 & M69I \\ 
11& TEM-51 & R244H \\
12& TEM-54 & R244L \\
13& TEM-55 & G218E \\
14& TEM-57 & G92D \\ 
15& TEM-70 & R204Q\\ 
16& TEM-76 & S130G\\ 
17& TEM-79 & R244G\\    
18& TEM-84 & N276D\\ 
19& TEM-90 & D115G\\ 
20& TEM-95 & P145A \\
21& TEM-96 & D163G \\
22& TEM-103 & R275L\\
23& TEM-104 & A280V\\
24& TEM-105 & S124N\\
25& TEM-117 & L21F\\
26& TEM-122 & R275Q\\
27& TEM-127 & H158N\\
28& TEM-128 & D157E\\
29& TEM-135 & M182T\\
30& TEM-141 & K34E\\
31& TEM-143 & R164C\\
32& TEM-148 & T189K\\
33& TEM-150 & E28D\\ 
34& TEM-156 & M155I\\
35& TEM-166 & R120G\\
36& TEM-168 & T265M\\
37& TEM-170 & G283C\\
38& TEM-171 & V84I\\
39& TEM-174 & A213V\\
40& TEM-176 & A224V\\
41& TEM-181 & A184V\\
42& TEM-183 & F230L\\
43& TEM-186 & D176N\\
44& TEM-191 & E240K\\
45& TEM-192 & M68I\\
46& TEM-198 & T271I\\
\hline
\end{tabular}
\end{table}

\begin{table}
\caption{The double mutants of the
record which combine single mutations
of the record}
\centering
\begin{tabular}{l l  l l}  
\hline \hline \\
1 &TEM-6: &E104K &R164H\\

2 &TEM-7: &Q39K &R164S\\

3 &TEM-10: &R164S &E240K\\

4 &TEM-11: &Q39K &R164H\\

5 &TEM-13: &Q39K &265M\\

6 &TEM-15: &E104K &G238S\\

7 &TEM-18: &Q39K &E104K\\

8 &TEM-20: &M182T &G238S\\

9 &TEM-26: &E104K &R164S \\

10 &TEM-28: &R164H &E240K\\

11 &TEM-32: &M69I &M182T\\

12 &TEM-35: &M69L &N276D\\

13 &TEM-36: &M69V &N276D\\

14 &TEM-37: &M69I &N276D\\

15 &TEM-38: &M69V  &R275L\\

16 &TEM-44: &Q39K &R244S\\

17 &TEM-45: &M69L &R275Q\\

18 &TEM-53: &L21F &R164S\\

19 &TEM-59: &Q39K &S130G\\

20 &TEM-65: &Q39K &R244C\\

21 &TEM-71: &G238S &E240K\\

22 &TEM-77: &M69L &R244S\\

23 &TEM-82: &M69V &R275Q\\

24 &TEM-106: &E104K &M182T\\

25 &TEM-110: &L21F &T265M \\

26 &TEM-115: &L21F &R164H\\

27 &TEM-116: &V84I  &A184V\\

28 &TEM-118: &R164H &T265M\\

29 &TEM-120: &L21F &G238S\\

30 &TEM-144: &R164C &E240K\\

31 &TEM-147: &R164H &A224V\\

32 &TEM-154: &M69L &R164S\\

33 &TEM-160: &Q39K &M69V\\

34 &TEM-165: &R164S &M182T\\

35 &TEM-189: &M69L &E240K\\
\hline
\end{tabular}
\end{table}

\newpage

Ceftazidime

$w(L21F),  w(R164S), w(E240K), w(T265M)  > w(\text{{TEM-1}})$

$w(\{R164S, E240K\})  > w(R164S), w(E240K)$

$w(\{L21F, R164S\})  > w(L21F)$

$w(\{L21F, T265M\})  > w(T265M)$

$w(\{R164S, T265M\}) > w(T265M)$

$w(\{E240K, T265M\})   > w(T265M)$

\smallskip
Cefotaxime

$w(L21F), w(R164S), w(E240K), w(T265M) > w(\text{{TEM-1}})$

$w(\{R164S, E240K \})> w(R164S), w(E240K)$

$w(\{L21F, R164S \})> w(L21F)$

\smallskip
Pipercillin/tazobactam penicillin/inhibitor:

$w(L21F), w(T265L) > w(\text{{TEM-1}})$

$w(\{L21F,T265M\}) > w(L21F), w(T265M)$

\smallskip
Cefpodoxime:

$w(R164S), w(E240K) > w(\text{{TEM-1}})$

$w(\{ R164S, E240K\})> w(E240K), w(R164S)$

\smallskip
Cefotetan:

$w(R164S), w(E240) > w(\text{{TEM-1}})$   

$w(\{ R164S, E240K \})> w(R164S),  w(E240)$

\smallskip
Cefprozil:

$w(L21F), w(T265M) > w(\text{{TEM-1}})$

$w(\{L21F, T265M\}) > w(L21F)$
  
\smallskip
Ampicillin:

$w(L21F), w(T265M) > w(\text{{TEM-1}})$

No double mutants to list.

\smallskip
Cefepime: 

$w(L21F), w(R164S) > w(\text{{TEM-1}})$   

No double mutants to list.

\smallskip
Amoxillin+Clavulanate:

$w(L21F), w(T265M) > w(\text{{TEM-1}})$

No double mutants to list.

\begin{table}
\begin{tabular}{c l}
\multicolumn{2}{c}
{Admino acids and possible substitutions}\\
\hline \hline \\
&R: C, G, H, I, K, L, M, P, Q, S, T, W\\
&S: A, C, F, G, I, L, N, P, T, Y, W\\
&L: F, H, I, M, P, Q, R, S, V, W\\
&I: F, K, L,  M, N, R, S, T, V\\
&G: A, C, D, E, R, S, V, W\\
&T: A, I, K, M, N, P, R, S\\ 
&Q: D, E, H, K, L, N, P, R\\ 
&V: A, D, E, F, G, I, L, M \\
&A: D, E, G, P, S, T, V\\
&P: A, H, L, Q, R, S, T\\
&D: A, E, G, H, N, V, Y\\
&H: D, L, N, P, Q, R, Y\\
&K: E, I, M, N, Q, R, T\\
&N: D, H, I, K, S, T, Y\\
&E: A, D, G, K, Q, V \\
&F: C, L, I, S, V, Y\\ 
&M: I, K, L, R, T, V\\ 
&Y: C, D, F, H, N, S \\
&C: F, G, R, S, W, Y\\
&W: C, G, L, R, S\\
\hline\\
\end{tabular}
\end{table}

\end{document}